\documentstyle[floats,aps,preprint]{revtex}
\begin{document}
\draft
\preprint{\vbox{ \hbox{SOGANG-HEP 255/99} \hbox{hep-th/9902093}  }}
\title{Absorption cross section and Hawking radiation
 in two-dimensional AdS black hole}
\author{Won Tae Kim\footnote{electronic address:wtkim@ccs.sogang.ac.kr}, 
        John J. Oh\footnote{electronic address:john5@cosmos.sogang.ac.kr}
   and  Jung Hee Park\footnote{electronic address:pjhee@cosmos.sogang.ac.kr}}
\address{Department of Physics and Basic Science Research Institute,\\
         Sogang University, C.P.O. Box 1142, Seoul 100-611, Korea}
\maketitle
\bigskip 
\begin{abstract}
We calculate the absorption coefficient of
scalar field 
on the background of the two-dimensional AdS black hole,
which is of relevance to Hawking radiation. 
For the massless scalar field, we find that there
does not exist any massless radiation.
\end{abstract}
\bigskip
\newpage
There has been much attention to the lower-dimensional black
holes to study the final state of black holes at the end
of evaporation \cite{cghs,rst} without encountering
some complexities of realistic four-dimensional gravity. 
The evaporation of the black hole is essentially based 
on Hawking radiation
\cite{haw}.
The Hawking radiation of Callan-Giddings-Harvey-Strominger(CGHS) 
model is given by the vacuum expectation value 
of energy-momentum tensors, which means that the black hole is
no more black, instead, grey hole after quantization. 
Therefore the absorption coefficient \cite{dm,ms} 
is less than one and
the reflection coefficient should be nontrivial.

On the other hand, two-dimensional anti-de Seitter(AdS$_2$)
geometry which has
a constant curvature scalar appears in the CGHS model
with the help of quantum back reaction of the geometry
on a constant dilaton background \cite{sol,kim},
where the AdS$_2$ black hole geometry is given by
\begin{equation}
  \label{metric}
  ds^2 = -\left(-M+\frac{r^2}{\ell^2}\right)
dt^2 + \left(-M+\frac{r^2}{\ell^2}\right)^{-1}dr^2
\end{equation}
where the horizon is 
$r_H=\sqrt{M}\ell$.
This metric (\ref{metric}) is asymptotically nonflat and
has a constant negative curvature scalar $R=-\frac{2}{\ell^2}$
with a negative cosmological constant $\Lambda=-\frac{1}{\ell^2}$, 
which is drastically different from the asymptotic flat CGHS model.
So one might wonder whether conventional Hawking radiation
appears or not in such a AdS$_2$ black hole. 
The recent study \cite{kim} shows that the AdS black hole
may be quantum-mechanically stable without Hawking radiation
as far as we consider massless radiation from the black
hole. In Ref. \cite{kim}, the energy-momentum tensor
for the N massless scalar fields 
on the AdS$_2$ black hole background yields
remarkably vanishing result as
\begin{equation}
\label{nhr}
T_{--}^{\rm Qt} (\sigma^+,\sigma^-)
= T_{--}^{\rm Bulk}+ T_{--}^{\rm boundary}= 0
\end{equation}
where the energy-momentum tensor induced by quantum correction $T_{--}^{Qt}$ is conveniently defined by bulk and boundary contributions,
\begin{eqnarray}
T_{--}^{\rm Bulk}(\sigma^+,\sigma^-)&=& -\frac{\kappa M}{4 \ell^2} ,\label{bulk} \\  
T_{--}^{\rm boundary}(\sigma^+,\sigma^-) 
&=& \frac{\kappa M}{4 \ell^2} 
\label{boundary}
\end{eqnarray} 
in the conformal gauge, $g_{+-} = -\frac{1}{2} e^{2\rho (\sigma^{+},
  \sigma^{-})}$, $g_{\pm\pm}=0$,and $\kappa = \frac{N}{12}$.
For the asymptotically flat CGHS black hole, 
however, the 
Hawking radiation has been determined by only the boundary contribution
since the bulk part vanishes at the spatial infinity.
In our AdS case, the bulk contribution is definite due to
the nontrivial asymptotic behavior of the
geometry. 

One might still wonder how does the
Hawking radiation of AdS$_2$ black hole vanish.
So, in this Brief Report, 
we shall study massless
scalar field as a test probe 
on the AdS$_2$ black hole background 
and calculate  
absorption(or reflection) coefficient, 
which will be a different test
whether Hawking radiation appears or not in AdS$_2$ black hole. 

Let us now consider the wave equation for the massless scalar
field on the AdS$_2$ black hole background (\ref{metric}), 
which is given by
\begin{equation}
  \label{eqmotion}
  \Box\Psi(t,r)=0
\end{equation}
with $\Psi(t,r)=e^{-i\omega t}\Phi(r)$,
and it yields explicitly
\begin{equation}
\label{radial}
\left(r^2-r_H^2 \right)
\partial_r^2 \Phi(r)+2r\partial_r \Phi(r)
+\frac{\omega^2 \ell^4}{r^2 -r_H^2}\Phi(r)=0.
\end{equation}
By changing variable 
$r$ in terms of $z=\frac{r-r_H}{r+r_H}$ ($0 \leq z \leq 1$),
Eq.(\ref{radial}) becomes
\begin{equation}
\label{singular}
 z(1-z) {\partial_z}^2\Phi(z)+
(1-z)\partial_z\Phi(z)+
\frac{\omega^2\ell^4}{4r_H^2}\left(\frac{1}{z}-1 \right) \Phi(z)=0,
\end{equation}
it is exactly solved as
\begin{equation}
  \label{general}
  \Phi(r) = C_1 \left( \frac{r-r_{H}}{r+r_{H}}\right)^{-\alpha} + C_2
  \left( \frac{r-r_{H}}{r+r_{H}} \right)^{\alpha},
\end{equation}
where $\alpha = \pm \frac{i \omega \ell^2}{2 r_{H}}$.
This solution (\ref{general})
contains the ingoing and outgoing wave modes together,
which will be easily seen by rewriting Eq.(\ref{general})
as 
\begin{equation}
  \label{near}
  \Phi(r)=C_1e^{-\frac{i\omega\ell^2}{2r_H}\ln(\frac{r-r_H}{r+r_H})}+C_2e^{\frac{i\omega\ell^2}{2r_H}\ln(\frac{r-r_H}{r+r_H})}.
\end{equation}
Thus the coefficient $C_1$ and $C_2$ are defined with ingoing and outgoing amplitudes,
respectively.  
The wave equation 
is exactly solved on the $AdS_{2}$ black hole
background, and we need not have matching procedure, 
which is contrasted to the higher-dimensional analysis of greybody
factors \cite{ms}.
We now choose boundary condition in the near horizon region 
and simply set $ C_2=0$, which means that the outgoing mode does
not exist in the near horizon limit \cite{ms}.
On the other hand, the solution in the far region naturally contains
only ingoing mode, 
and modes mixing does not occur. 
Once it means that ingoing mode starts at the far region, it does not
change its direction in the massless case.
This feature is drastically different from that of the wave solution of
the three-dimensional AdS black hole so called 
Ba$\tilde{{\rm n}}$ados-Teitelboim-Zanelli(BTZ)
black hole \cite{btz} in that the mode mixing certainly appears 
and it is responsible for the 
Hawking radiation. 

Now, the flux is defined by the
conserved current as
\begin{equation}
  \label{abs1}
F=\frac{2\pi}{i}\left(\frac{r^2-r_H^2}{\ell^2}\right)
(\Phi^*\partial_r\Phi-\Phi\partial_r\Phi^*),  
\end{equation}
and it is simply given by
\begin{equation}
  \label{flgen}
F= - 4 \pi \omega | C_1 |^2,
\end{equation}
by the use of Eq.(\ref{near}).
As easily seen, the flux does not depend on the
outgoing amplitude everywhere outside the horizon, and the  
absorption coefficient defined by
\begin{equation}
  \label{abscodef}
  A=\frac{F_{{\rm near}}}{F_{{\rm far}}},
\end{equation}
is exactly given as
\begin{equation}
\label{absco}
  A=1.
\end{equation}
From Eq.(\ref{absco}), we find our black hole
is a (perfect) black body rather than the grey hole. 
The physical meaning of this result is remarkable in
that there is no massless radiation in the AdS$_2$ spacetime.
This result is already shown in Ref.\cite{kim} by the use of the different
method, direct calculation of stress-tensor on the AdS$_2$ black hole. 

Finally we compare our result with three-dimensional calculation of 
absorption coefficient of BTZ black hole in Ref. \cite{bss}.
For a limiting case of vanishing azimuthal angular momentum of scalar field 
in the nonrotating BTZ black hole, there is still Hawking radiation,
which is contrast with the AdS$2$ black hole case.
This is because the radial equation in BTZ black hole
is different from Eq. (\ref{radial}), and
the reflection coefficient is nonzero.
In three-dimensions, in fact, ingoing mode in the near horizon 
is decomposed into ingoing and outgoing modes in the
far region, and this is essential reason why the Hawking
radiation appears. However, in our AdS$_2$ case, 
the ingoing mode in the near horizon is still ingoing mode
in the far region, which reflects no massless 
Hawking radiation in AdS$_2$ black hole.         
\vspace{20mm}

{\bf Acknowledgments}\\
W.T. Kim would like to thank Mariano Cadoni for helpful discussion on 
Hawking radiation and Sergey N. Solodukhin for bringing his papers related
to AdS$_2$ to my attention and Seungjoon Hyun and Julian Lee for
enlightening discussion and hospitality 
during the course of visiting Korea Institute
of Advanced Study. 
We also thank Myungseok Yoon for valuable comments.
This work was supported by Korea Research Foundation, 
No. BSRI-1998-015-D00074. 


\end{document}